\documentclass{article}

\usepackage[preprint]{neurips_2026}
\usepackage[utf8]{inputenc}
\usepackage[T1]{fontenc}
\usepackage{hyperref}
\usepackage{url}
\usepackage{booktabs}
\usepackage{amsmath,amssymb,amsthm}
\usepackage{graphicx}
\usepackage{subcaption}

\newtheorem{proposition}{Proposition}

\title{Governance Outruns Analysis:\\Incidence and Manipulability of Credential-Gated Reciprocal Review}

\author{%
  Jinming Xing \\
  North Carolina State University \\
  \texttt{jxing6@ncsu.edu} \\
  \And
  Charlotte Brian \\
  Independent Researcher \\
  \texttt{immjomrtal@gmail.com} \\
}

\begin{document}

\maketitle

\begin{abstract}
Generative tools have lowered the cost of producing a submission without lowering
the cost of evaluating one, and venues have responded with governance: submission
caps, reviewing obligations tied to submission count, and eligibility gates keyed
to prior publication. These rules bind tens of thousands of researchers and are
typically announced without published analysis of their incentive properties. We
formalise the family as credential-gated reciprocal review and prove three
structural results: a step-shaped obligation against linear demand yields
incidence that is not monotone in consumption; a transferable credential gate
defeats padding-resistant quota design, since the gate confers feasibility and no
per-author price offsets a jump in the feasible set; and a per-author obligation
with a per-paper penalty makes sanction risk perfectly correlated within an
author's portfolio. Evaluating against a full prior-cycle submission graph
(19{,}814 submissions, 64{,}648 authors, 75{,}859 observed reviews), the
obligation covers 84.2\% of realised demand, 81.4\% of authors owe nothing, and
the most burdened agents owe 4.45 times the review labour they generate against
0.45 times for the least. A three-part modification, evaluated on the same fixed
graph, flattens incidence to near-uniform and removes the padding incentive.
\end{abstract}

\section{Why this is an AI-native governance problem}

Two quantities have moved in opposite directions. The cost of producing a
submission has fallen, and the cost of evaluating one has not. Across two
consecutive cycles of the venue we study, submissions grew 69.8\% (11{,}672 to
19{,}814) while total reviews written grew 62.3\% (46{,}748 to 75{,}859). The gap
appears as dilution: reviews per paper fell from 4.005 to 3.829. Part of the
growth was absorbed by reviewing each paper slightly less.

The same pressure has reached the review side, where evidence that reviews are
themselves generated now supports benchmark datasets and detector evaluations
\citep{rao2025detecting,yu2025reviewed,demetrio2025genreview}. Governance is
therefore asked to solve two coupled problems: too many submissions per reviewer,
and reviews whose informational content cannot be assumed. The deployed response
is to ration submissions and conscript reviewers from the author pool, coupling
the right to submit to the duty to review and gating who may discharge it on
prior publication. More radical restructurings have been proposed
\citep{sankaralingam2025impact}, but quota rules are what venues have shipped, and
they are announced between cycles, bind immediately, and are copied if they
appear to work.

This paper asks what happens when academic infrastructure becomes AI-native
before its governance is ready. Our answer for one slice of it is that governance
is being written faster than it is analysed, and that the analysis is tractable
from public data before the rules take effect. The authorship-side result bears
directly on credit integrity: the gate makes a credentialed co-author a necessary
input to a submission, and no quota rule can price the resulting incentive to add
one.

\section{The mechanism family}

A \emph{credential-gated reciprocal review} mechanism is a tuple
$(\bar K, \gamma, \rho, r, \pi, \tau_f, X)$: $\bar K$ caps submissions per author,
$\gamma$ is an eligibility predicate fixed exogenously and not producible within a
cycle, $\rho$ requires each submission to carry a registered author satisfying
$\gamma$, $r$ maps an author's paper count to a review obligation, $\pi$ penalises
non-completion, $\tau_f$ freezes the author set, and $X$ resolves cap violations.

Write $k_i$ for papers per author, $n_p$ for authors per paper, $\bar n$ for its
mean, $\lambda$ for reviews per paper. Demand is $D = \lambda m$ over $m$
submissions, with $\sum_i k_i = \sum_p n_p = m\bar n$. Since a paper's demand
belongs to the paper rather than to each author on it, demand per author-slot is
$c := \lambda/\bar n$ and net contribution is $\nu(k_i) := r(k_i) - c k_i$, giving
$\sum_i \nu(k_i) = S - D$. Charging each author the full $\lambda$ per paper
overstates demand by a factor of $\bar n$; \citet{shah2026harmonic} makes the
corresponding point on the submission-cap side.

The deployed instance \citep{iclr2027author,iclr2027reviewer} sets $\bar K = 20$;
$\gamma$ as one accepted publication at any of 24 listed venues; $\rho$ as one
qualified author registered for at least 3 reviews, with teams containing no
qualified author exempt but capped at one submission; $r(k) = 6$ for $k \geq 3$;
$\pi$ as withheld access to one's own reviews plus desk rejection for $k \geq 3$;
$X$ as uniform random desk rejection.

\section{Three structural results}

\begin{proposition}[Two-step schedule]
\label{prop:steps}
A registered author's obligation is
$r(k) = 3\cdot\mathbf{1}\{k \geq 1\} + 3\cdot\mathbf{1}\{k \geq 3\}$ on
$1 \leq k \leq \bar K$, so $\Delta r(k) = 0$ for all
$k \in \{2\} \cup \{4,\dots,20\}$.
\end{proposition}

Seventeen of twenty permitted submissions carry zero marginal review cost, since
an author with $k \geq 3$ already reviews $6 \geq 3$ and so discharges $\rho$ for
every paper they appear on. The threshold at $k = 3$ is a notch in the sense of
\citet{kleven2013notches}: with utility $b_i k - c_r r(k)$ and $\Delta r = 0$ above
3, no interior optimum exists there, so agents bunch at $k = 2$ or jump past and
$k = 3$ is chosen by a measure-zero set. It is a compound threshold, since the
penalty escalates at the same point.

\begin{proposition}[Gate impossibility]
\label{prop:gate}
Let $\varphi(a)$ be the lead author's credit share on an $a$-author paper,
non-increasing with $\varphi(2) > 0$, and let the gate cap ungated output at
$\bar q$. If $\bar K > \bar q\,\varphi(1)/\varphi(2)$, then no submission quota
rule $f$ that charges collaborative work less than solo work is resistant to
author-list padding.
\end{proposition}

An unqualified lead author places $\bar q$ papers, earning $\bar q\varphi(1)$; one
qualified co-author makes $M := \min(m, \kappa/f(2), \bar K)$ feasible, earning
$M\varphi(2)$, so padding pays iff $M > \bar q\varphi(1)/\varphi(2)$. Blocking it
requires $\kappa/f(2) \leq \bar q\varphi(1)/\varphi(2)$, but $f(2) \leq f(1)$ gives
$\kappa/f(2) \geq N_1$, the solo allowance. The credit model
\citep{hodge1981publication,hagen2008harmonic,hagen2013harmonic} enters only through
$\varphi(1)/\varphi(2)$ while the return scales with $\bar K$, so the result needs
dilution bounded, not of a particular form: at the deployed parameters padding pays
unless a second author strips 95\% of the lead author's credit. The reason is
structural. Quota rules price submissions, the gate determines feasibility, and no
per-author price offsets a jump in the feasible set, so the gate defeats the
padding-resistance sought by \citet{shah2026harmonic} and
\citet{mumcu2026roleaware} alike.

\begin{proposition}[Correlated failure]
\label{prop:risk}
If the obligation is discharged per author but the penalty applies to each of the
author's submissions, all $k$ papers share one completion event. With failure
probability $p$, expected sanctions are $pk$ under both independent and
correlated failure, but the variance is $kp(1-p)$ versus $k^2p(1-p)$.
\end{proposition}

The mechanism maximises the variance of desk rejections for a given mean, and
exposure concentrates on papers whose only qualified author has high $k$, which by
the ungated cap are exactly the papers that had to recruit a credentialed
co-author. That team commits at $\tau_f$ and can neither observe completion effort,
change the author set afterwards, nor diversify.

\section{Measurement}

We apply the rules to the complete prior-cycle submission graph: 19{,}814
submissions, 64{,}648 authors, $\bar n = 5.682$. Demand is measured rather than
assumed, by counting review notes: 75{,}859 reviews, $\lambda = 3.829$. Enforcing
that the registrant be qualified, minimal compliant supply is $S = 63{,}879$, so
$S - D = -11{,}980$ and the obligation covers \textbf{84.2\%} of realised demand.

\begin{figure}[htbp]
  \centering
  \begin{subfigure}{0.48\textwidth}
    \includegraphics[width=\linewidth]{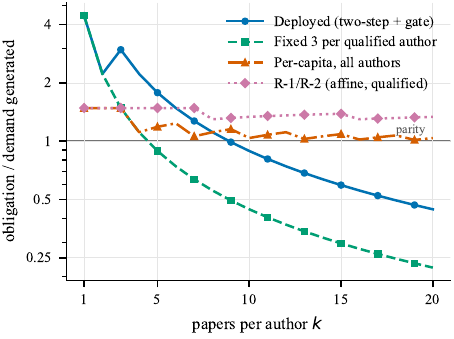}
    \caption{Obligation divided by demand generated.}
    \label{fig:incidence}
  \end{subfigure}\hfill
  \begin{subfigure}{0.48\textwidth}
    \includegraphics[width=\linewidth]{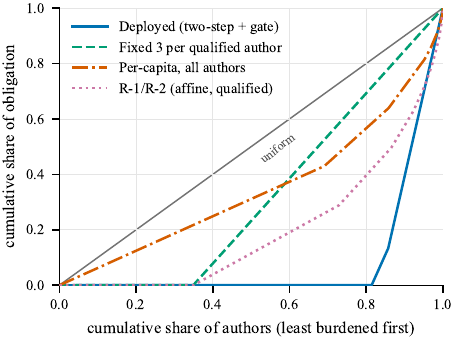}
    \caption{Concentration across authors.}
    \label{fig:lorenz}
  \end{subfigure}
  \caption{The deployed schedule is the only rule considered whose incidence is
  non-monotone, and the only one with a notch. Appendix~\ref{app:extra} gives the
  distribution it is levied on and the sensitivity to $\lambda$.}
  \label{fig:main}
\end{figure}

Figure~\ref{fig:main} shows the two findings independent of $\lambda$. Incidence
spans a factor of ten and is not monotone: the most burdened agents owe 4.45 times
the review labour they generate, the least burdened 0.45 times, and two authors
both owe 6 reviews when one submits 3 papers and the other 20. The obligation is
also concentrated, with 81.4\% of authors owing nothing and the top fifth owing
essentially all of it.

Attributing the gap by tier reverses a natural intuition: the $k \geq 3$
population, 14.3\% of authors, is a net contributor of roughly $+24{,}000$ reviews
while the exempt majority accounts for the entire shortfall, and agents above the
net-contribution threshold $k^\star = 6\bar n/\lambda \approx 8.9$ account for under
7\% of the gross deficit. The high-volume submitters the obligation targets are the
ones subsidising it.

\begin{table}[htbp]
\caption{Credential breadth on nested venue lists. Exposed teams are those for
whom acquiring a credentialed co-author lifts output from $\bar q$ to $\bar K$.
Computed over venues hosted on the review platform, so the comparison is relative.}
\label{tab:breadth}
\centering
\small
\begin{tabular}{lrrrr}
\toprule
Credential definition & Qualify & Ungated & Exposed teams & Sole-qualified \\
\midrule
Full published list    & 64.3\% & 9.2\%  & 5{,}558  & 9.2\% \\
ICLR/NeurIPS/ICML/TMLR & 64.0\% & 9.4\%  & 5{,}668  & 9.3\% \\
ICLR/NeurIPS/ICML      & 62.5\% & 10.1\% & 6{,}098  & 9.7\% \\
ICLR only              & 45.4\% & 18.3\% & 12{,}068 & 13.6\% \\
\bottomrule
\end{tabular}
\end{table}

The gate is slack. 65.1\% of authors satisfy $\gamma$, only 8.7\% of submissions
carry no qualified author, and the most papers any author carries as sole
credential-holder is 6, far below the licensing capacity of 20 the cap permits,
because $k$ is bounded by productivity rather than by the rule. Tightening does
not help. Table~\ref{tab:breadth} varies the credential over nested venue lists:
the published 24-venue list moves qualification by 0.3 points relative to ICLR,
NeurIPS, ICML and TMLR alone, because authors submitting here overwhelmingly hold
one of the core four already, and cutting to ICLR only screens harder while
doubling the population facing an unchanged padding return of
$\bar K/\bar q = 20$. Screening and exposure move together, and only the latter
is a mechanism-design cost. Correlated exposure covers 899 submissions whose sole
qualified author sits in the desk-rejection tier.

\section{A targeted modification}

\textbf{R-1}: replace the step schedule with $r(k) = c_\gamma k$ rounded by largest
remainder, $c_\gamma$ being demand per qualified author-slot, so supply matches
demand exactly and the notch disappears. \textbf{R-2}: let the credential confer
obligation rather than permission, so any team may submit up to $\bar K$ and a
qualified co-author adds review duty rather than allowance; the feasibility jump
driving Proposition~\ref{prop:gate} vanishes and padding has zero return, while the
gate still keeps unqualified authors out of the reviewer pool. \textbf{R-3}: replace
per-author desk rejection with carry-forward debt against the next cycle's
allowance, restoring independence across a portfolio. Table~\ref{tab:exp2} holds the
observed graph fixed and varies only the rule, so the comparison involves no
behavioural model and no fitted parameters.

\begin{table}[htbp]
\caption{Mechanism comparison on the fixed graph. Demand is the measured
75{,}859 reviews.}
\label{tab:exp2}
\centering
\small
\begin{tabular}{lrrrl}
\toprule
Rule & Demand covered & Owe zero & Notch & Incidence $k{=}1 \to k{=}20$ \\
\midrule
Deployed (two-step + gate)   & 84.2\%  & 81.4\% & at $k{=}3$ & 4.45$\times \to$ 0.45$\times$ \\
Fixed 3 per qualified author & 166.5\% & 34.9\% & none & 4.45$\times \to$ 0.22$\times$ \\
Per-capita, all authors      & 137.2\% & 0.0\%  & none & 1.48$\times \to$ 1.04$\times$ \\
\textbf{R-1/R-2}             & \textbf{100.0\%} & 34.9\% & none & \textbf{1.48$\times \to$ 1.34$\times$} \\
\bottomrule
\end{tabular}
\end{table}

The modification flattens incidence from a tenfold spread to near-uniform while
keeping unqualified authors out of the reviewer pool, and makes explicit a cost the
deployed rule conceals: exempting 34.9\% of authors means the rest pay about 1.5
times their own consumption. The deployed schedule pays the same bill by charging
some agents 4.45 times and others 0.45 times. We claim no capacity gain, since the
deployed rule already covers 84.2\%; the gains are in incidence,
manipulation-resistance and risk structure.

\section{Related work}

Peer review market design has treated reviewer eligibility as universal or
effort-determined. \citet{fernandes2025market} propose Admission Control, rejecting
submissions of reviewers who fail to exert effort, attaining first-best welfare
under observable effort. The mechanism we study deploys something close to it at
scale, then adds a gate keyed to an exogenous credential; because that credential is
transferable through co-authorship and not producible within a cycle, it lies
outside the action space their model admits, and their assumption that every agent
is both author and reviewer is what the gate breaks. On the submission side,
\citet{shah2026harmonic} derives a harmonic quota whose cost falls with co-author
count and \citet{mumcu2026roleaware} extend it with author roles. Both design the
submission price; Proposition~\ref{prop:steps} concerns the reviewing obligation,
the other half of the same ledger, and Proposition~\ref{prop:gate} shows the gate
defeats the padding-resistance both seek. \citet{song2025coauthor} study the
author's best response under a predecessor nomination policy, and
\citet{zhang2023system} analyse review at the system level.

\section{Conclusion}

Quota and reciprocal-review rules are the first governance response to AI-driven
submission growth that binds every author at a major venue, and they were deployed
without published analysis. That analysis turns out to be cheap: three properties
follow from the shape of the rules alone, and the quantities that matter are
recoverable from public records before a single paper is submitted under them. The
obligation is mis-calibrated in incidence rather than in aggregate, the credential
gate screens almost nobody while creating a padding incentive no quota rule can
price, and tying a per-author duty to a per-paper penalty concentrates correlated
risk on the teams the exemption was written to protect; each is repairable by
changing one component, and the repairs compose because the components are
independent. The broader point is procedural. Our bunching prediction is stated
before the mechanism has run, against a distribution we have measured, and it will
be checkable against the realised distribution afterwards. Venue rules are
unusually well suited to this discipline, since the mechanism is public, the
population is enumerable, and the counterfactual is one cycle old, so analysing
rules before they bind rather than after they entrench is achievable at the speed
governance is actually moving.

\bibliographystyle{plainnat}
\bibliography{refs}

\appendix

\section{Proofs}
\label{app:proofs}

\paragraph{Proposition~\ref{prop:steps}.} The obligation has two sources. The
reciprocal requirement binds every author on three or more papers, imposing 6
reviews. The registration requirement binds each submission, requiring one
qualified author registered for at least 3. An author with $k \geq 3$ already
reviews $6 \geq 3$ and so discharges registration for every paper they appear on,
so the requirement binds only on papers all of whose qualified authors have
$k \leq 2$, where exactly one registers. A registered author's obligation is thus
3 for $k \in \{1,2\}$ and 6 for $k \geq 3$. Taking first differences gives
$\Delta r$, zero except at $k = 1$ and $k = 3$.

\paragraph{Net-contribution threshold.} On $k \geq 3$, $\nu(k) = 6 - ck$ is
strictly decreasing with unique zero $k^\star = 6/c = 6\bar n/\lambda$. On
$k \in \{1,2\}$, $\nu(k) = 3 - ck > 0$ whenever $c < 3/2$, which holds at the
measured $c = 0.674$. That $\sum_i \nu(k_i) = S - D$ follows from
$\sum_i c k_i = (\lambda/\bar n)(m\bar n) = D$.

\paragraph{Bunching.} With $\Delta r(k) = 0$ on $\{4,\dots,\bar K\}$, $U$ is
linear and increasing there, so any agent choosing $k \geq 3$ chooses the largest
feasible $k$ and it suffices to compare $k = 2$ against $k = \bar K$:
$U(\bar K) - U(2) = b_i(\bar K - 2) - 3c_r$, giving $b^\star = 3c_r/(\bar K - 2)$.
Agents below $b^\star$ locate at $k = 2$, agents above at $\bar K$ or their
exogenous project-supply limit. The set choosing exactly $k = 3$ has measure zero
under any atomless distribution of $b_i$. Since the project-supply limit binds
before $\bar K$ for almost every agent, predicted mass lands across
$\{3, 4, \dots\}$ rather than at $\bar K$; the sign of the prediction at $k = 2$
and $k = 3$ is unaffected.

\paragraph{Proposition~\ref{prop:gate}.} Given in the main text.

\paragraph{Proposition~\ref{prop:risk}.} Let $F$ indicate that an author fails to
complete, $\Pr[F = 1] = p$. Under the per-paper penalty the number of sanctioned
papers is $kF$, so $\mathbb{E}[kF] = pk$ and $\mathrm{Var}[kF] = k^2p(1-p)$. Under
independent per-paper failure at the same marginal rate the count is
$\sum_{j \leq k} F_j$, with mean $pk$ and variance $kp(1-p)$.

\section{Additional results}
\label{app:extra}

\begin{figure}[htbp]
  \centering
  \begin{subfigure}{0.48\textwidth}
    \includegraphics[width=\linewidth]{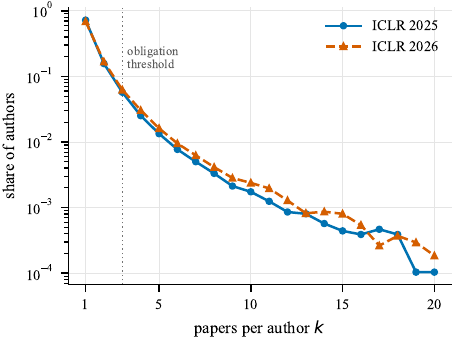}
    \caption{Papers per author, two consecutive cycles.}
    \label{fig:kdist}
  \end{subfigure}\hfill
  \begin{subfigure}{0.48\textwidth}
    \includegraphics[width=\linewidth]{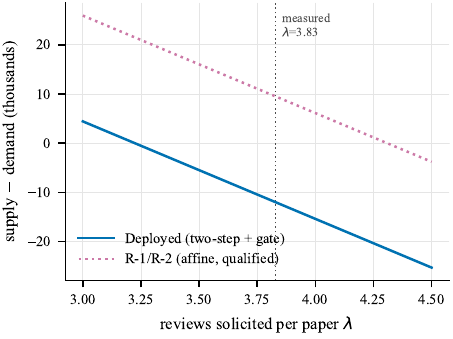}
    \caption{Aggregate gap as a function of $\lambda$.}
    \label{fig:lambda}
  \end{subfigure}
  \caption{Left: the distribution the obligation is levied on is stable across
  cycles, with the threshold marked. Right: the deployed rule crosses zero inside
  the plausible range of $\lambda$, so its aggregate adequacy depends on a
  parameter the venue sets after the rule is fixed; the modification tracks demand
  by construction at any $\lambda$.}
  \label{fig:dist}
\end{figure}

\paragraph{Calibration of the affine rule.} At $c_\gamma = 0.873$ reviews per
qualified author-slot, rounding each obligation up oversupplies by 12.6\% (85{,}383
reviews against demand of 75{,}859). Allocating the integer remainder by largest
fractional part yields 75{,}859, matching measured demand exactly. Table~\ref{tab:exp2}
reports the latter.

\paragraph{Supply decomposition.} Under the deployed rule, the 9{,}268 authors with
$k \geq 3$ owe 55{,}608 reviews and are net contributors of $+24{,}355$; the
4{,}049 low-tier registrants owe 12{,}147 and contribute $+8{,}791$; the 51{,}331
authors with no obligation account for $-41{,}250$. Enforcing that the registrant
be qualified reduces supply from 67{,}755 to 63{,}879 and widens the gap from
$-8{,}104$ to $-11{,}980$.

\paragraph{Submission cap.} 101 authors exceed $\bar K = 20$, with 721 author-slots
in excess and a maximum of 54 submissions by a single author. Under $X$ these
resolve by uniform random desk rejection.

\paragraph{Correlated exposure.} 1{,}811 submissions (9.1\%) have exactly one
qualified author; for 899 that author is in the desk-rejection tier. Exposure by
the carrier's $k$: 291 at $k = 3$, 197 at $k = 4$, 127 at $k = 5$, tailing to 26 at
$k = 10$. Proposition~\ref{prop:risk} binds more broadly, on every author with
$k \geq 3$; these 899 are the subset where a third party bears the risk.

\section{Data, limitations, and reproduction}
\label{app:data}

All quantities derive from public submission and review records for the cycle
preceding the mechanism's introduction. We recover author lists for every
submission, count review notes per submission to obtain $\lambda$ without assuming
a target, and assemble the qualified-author set by unioning author identifiers over
accepted papers at qualifying venues hosted on the same platform. The registration
requirement is solved as a minimum set cover by the standard greedy algorithm;
since each low-tier author covers at most two papers the gap to optimal is small,
and an exact minimum would yield fewer registrants and a larger measured shortfall.
Withdrawn and desk-rejected submissions are retained, since both consume quota and
empirically most consume reviews (means of 3.80 and 2.91). All results are CPU-only
and reproduce in a few minutes on a laptop; no accelerators are used, and the only
externally set quantity, $\lambda$, is measured rather than chosen.

The qualified-author set is a lower bound. Several listed qualifying venues are
hosted elsewhere, and for some older venues author identifiers are recorded as
email addresses rather than profile handles and cannot be joined, so authors whose
only qualifying publication predates roughly 2021 are under-counted. This biases
the count of ungated papers upward, which strengthens rather than weakens the
finding that the gate is slack. Extending the credential join to bibliographic
databases, and replicating on other venues that have adopted comparable rules
\citep{song2025coauthor,zhang2023system,fernandes2025market}, is the natural next
step.

The counterfactual applies the rules to a pool that was not optimising against
them, so it measures initial incidence rather than equilibrium response. Minimal
compliance is assumed throughout, which is the assumption most favourable to the
mechanism, so the measured shortfall is a lower bound. We also assume every
registered reviewer is activated; where a venue activates only a subset, realised
supply falls further.

\end{document}